\documentclass[prb,reprint,superscriptaddress]{revtex4-1}

\usepackage{graphicx}
\usepackage{dcolumn}
\usepackage{bm}
\usepackage{amsthm}
\usepackage{amsmath}
\usepackage{amssymb}
\usepackage{xcolor}

\begin{document}

\title{Anisometric mesoscale nuclear and magnetic texture in sintered Nd--Fe--B magnets}

\author{Ivan Titov}\email[Electronic address: ]{ivan.titov@uni.lu}
\author{Dirk Honecker}
\affiliation{Department of Physics and Materials Science, University of Luxembourg, 162A~Avenue de la Fa\"iencerie, L-1511 Luxembourg, Grand Duchy of Luxembourg}
\author{Denis Mettus}
\affiliation{Physik-Department, Technische Universit\"at M\"unchen, James-Franck-Stra{\ss}e, D-85748 Garching, Germany}
\author{Artem Feoktystov}
\affiliation{Forschungszentrum J\"ulich GmbH, J\"ulich Centre for Neutron Science (JCNS) at Heinz Maier-Leibnitz Zentrum (MLZ), Lichtenbergstra{\ss}e~1, D-85748, Garching, Germany}
\author{Joachim Kohlbrecher}
\affiliation{Paul Scherrer Institute, CH-5232~Villigen PSI, Switzerland}
\author{Pavel Strunz}
\affiliation{Nuclear Physics Institute, Department of Neutron Physics, CZ-25068 $\breve{R}$e$\breve{z}$, Czech Republic}
\author{Andreas Michels}\email[Electronic address: ]{andreas.michels@uni.lu}
\affiliation{Department of Physics and Materials Science, University of Luxembourg, 162A~Avenue de la Fa\"iencerie, L-1511 Luxembourg, Grand Duchy of Luxembourg}

\begin{abstract}
By means of temperature and wavelength-dependent small-angle neutron scattering (SANS) experiments on sintered isotropic and textured Nd-Fe-B magnets we provide evidence for the existence of an anisometric structure in the microstructure of the textured magnets. This conclusion is reached by observing a characteristic cross-shaped angular anisotropy in the total unpolarized SANS cross section at temperatures well above the Curie temperature. Comparison of the experimental SANS data to a microstructural model based on the superquadrics form factor allows us to estimate the shape and lower bounds for the size of the structure. Subtraction of the scattering cross section in the paramagnetic regime from data taken at room temperature provides the magnetic SANS cross section. Surprisingly, the anisotropy of the magnetic scattering is very similar to the nuclear SANS signal, suggesting that the nuclear structure is decorated by the magnetic moments via spin-orbit coupling. Based on the computation of the two-dimensional correlation function we estimate lower bounds for the longitudinal and transversal magnetic correlation lengths.
\end{abstract}

\maketitle

\section{Introduction}

Sintered Nd-Fe-B-based permanent magnets are multiphase materials, which find widespread technological application in electromotors, wind turbines, and in various consumer-electronics devices~\cite{gutfleisch2011}. The microstructure of these materials consists of micron-sized Nd-Fe-B crystallites (typical average grain size:~$\sim 5 \, \mu\mathrm{m}$) and various so-called Nd-rich phases~\cite{hono2012,woodcock2012}. As revealed by numerous electron-microscopy and atom-probe-tomography investigations (e.g., \cite{hono2009}), the Nd-rich phases form thin (several nm thick) layers at the $\mathrm{Nd}_2\mathrm{Fe}_{14}\mathrm{B}$ grain boundaries and they exist as well in the form of larger grains at the $\mathrm{Nd}_2\mathrm{Fe}_{14}\mathrm{B}$ grain junctions. The properties of the Nd-rich phases range from paramagnetic to ferromagnetic, metallic to oxide, or from crystalline to amorphous (see Table~1 in~\cite{woodcock2012}). They decisively determine the magnetic properties of Nd-Fe-B magnets.

Industrial grade Nd-Fe-B magnets are produced by the powder-metallurgical route. The final magnet is highly anisotropic (magnetically and structurally textured) due to the magnetic field applied during the synthesis process~\cite{gutfleisch2000}. The associated crystallographic texture shows up in x-ray wide-angle diffraction data, and the alignment of the easy axes of magnetization is observed in bulk magnetization measurements~\cite{liu2010,perigo2015,perigo2016no1}. However, whereas in hot-deformed Nd-Fe-B-based nanocomposites platelet-shaped precipitates with preferential orientation have been observed (e.g., \cite{SepehriAmin2013,michelspra2017}), we are not aware that similar \textit{anisometric} structures have been reported for sintered Nd-Fe-B on a mesoscopic length scale ($\sim 1-300 \, \mathrm{nm}$).

In this communication we report the results of a comparative small-angle neutron scattering (SANS) study of the microstructure of a sintered textured and isotropic Nd-Fe-B specimen. The SANS technique (see~\cite{rmp2019} for a recent review) provides statistically-averaged bulk information about both the structural and magnetic correlations on a length scale between a few and a few hundred of nanometers. This method has previously been applied to study the structures of magnetic nanoparticles~\cite{disch2012,guenther2014,bender2015,bender2018jpcc,bender2018prb,oberdick2018,krycka2019,benderapl2019,bersweiler2019}, soft magnetic nanocomposites~\cite{suzuki2007,herr08pss}, proton domains~\cite{michels06a,aswal08nim,noda2016}, magnetic steels~\cite{bischof07,bergner2013,Pareja:15,gilbert2016,shu2018}, or Heusler-type alloys~\cite{bhatti2012,runov2006,michelsheusler2019,leighton2019,sarkar2020}. Regarding Nd-Fe-B, SANS investigations~\cite{perigo2016no1,perigo2018} have reported a peculiar cross-shaped angular anisotropy in the scattering cross section of the polycrystalline textured material. In the single-crystal study by \textcite{kreyssig09} the cross-shaped anisotropy was interpreted in terms of a fractal-like magnetic domain structure.

These interesting results have motivated the present comparative SANS investigation. Here, we unambiguously demonstrate that the cross-shaped feature is of nuclear (i.e., nonmagnetic) origin. Our SANS analysis suggests the presence of anisometric structures in sintered textured Nd-Fe-B on a mesoscopic length scale. Further analysis of the purely magnetic SANS cross section at room temperature reveals that the nuclear texture becomes decorated by the magnetic spin structure, presumably via the spin-orbit coupling.

\begin{figure*}[tb!]
\centering
\includegraphics[width=1.0\linewidth]{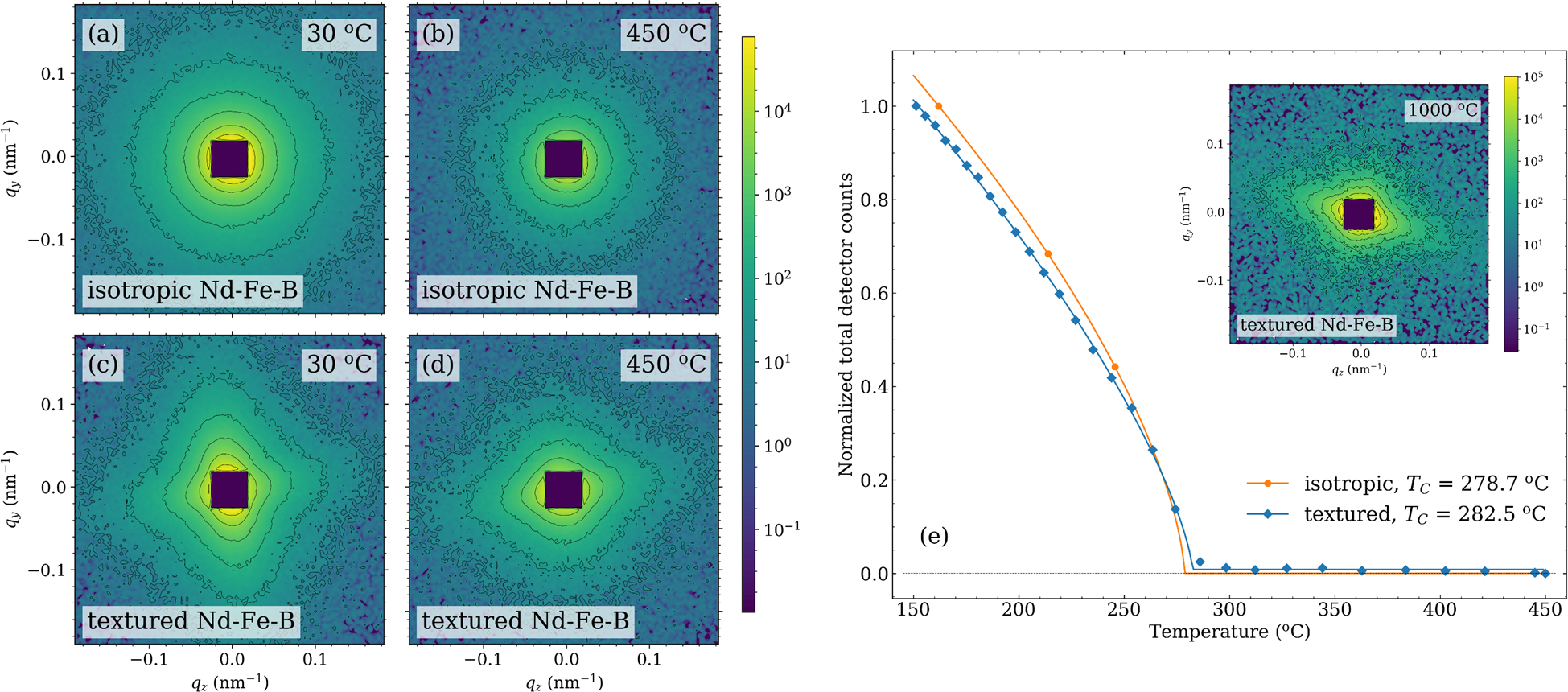}
\caption{\label{fig1} (a)--(d)~Total SANS cross section $d \Sigma / d \Omega$ of sintered isotropic and sintered textured Nd-Fe-B at several temperatures (see insets) (logarithmic color scale). (e)~Temperature dependence of the normalized integrated total neutron intensity of sintered isotropic and textured Nd-Fe-B (KWS-1, JCNS). The data have been normalized according to $(I - I_{\mathrm{min}})/(I_{\mathrm{max}} - I_{\mathrm{min}})$, where $I_{\mathrm{min}}$ and $I_{\mathrm{max}}$ denote, respectively, the minimum and maximum values of the integrated scattering intensities in dependence on temperature, separately for each sample. Error bars are of the order of the data-point size. Solid lines:~Fit of the data to $I_0 t^{\beta}$, where $I_0$, $\beta$, and $T_C$ are adjustable parameters, and $t = (T - T_C)/T_C$. The estimated Curie temperatures of the isotropic and textured Nd-Fe-B magnets are, respectively, $T_C = 278.7 \pm 0.1 \, ^{\circ}\mathrm{C}$ and $T_C = 282.5 \pm 0.8 \, ^{\circ}\mathrm{C}$. Note that $T_C = 312 \, ^{\circ}\mathrm{C}$ for single-crystalline $\mathrm{Nd}_2\mathrm{Fe}_{14}\mathrm{B}$~\cite{gutfleisch2000}. The inset in (e) shows the total $d \Sigma / d \Omega$ of the sintered textured Nd-Fe-B sample at $T = 1000 \, ^{\circ}\mathrm{C}$ (logarithmic color scale). The texture axis in the two-dimensional scattering maps is along the $q_z$-direction.}
\end{figure*}

\section{Experimental}
\label{exp}

Commercially available isotropic (grade:~N42) and textured (grade:~N38) Nd-Fe-B-based sintered magnets were used in this study. Temperature-dependent SANS measurements, between $30-1000 \, ^{\circ}\mathrm{C}$, were performed at the KWS-1 instrument~\cite{feoktystov2015} at the J\"ulich Centre for Neutron Science, Heinz Maier-Leibnitz Zentrum, Garching, Germany. We used unpolarized incident neutrons with a mean wavelength of $\lambda = 5.0 \, \mathrm{\AA}$ and a bandwidth of $\Delta\lambda/\lambda = 10 \, \%$ (FWHM). Wavelength-dependent ($\lambda = 5-18 \, \mathrm{\AA}$) SANS measurements at room temperature were carried out at SANS-1 at the Paul Scherrer Institute, Switzerland~\cite{aswal08nim}. SANS data were recorded for three sample-to-detector distances on the two Nd-Fe-B samples (isotropic and textured). The accessible range of momentum transfers in the experiment is about $0.02 \, \mathrm{nm}^{-1} \lesssim q \lesssim 1 \, \mathrm{nm}^{-1}$, so that real-space structure on a scale of roughly $1-300 \, \mathrm{nm}$ is probed. The direction of the texture axis of the anisotropic sample was perpendicular to the incident neutron beam and along the horizontal ($q_z$) direction. Prior to the temperature-dependent neutron measurement each sample was magnetized (externally, outside of the SANS setup) by a horizontal $9 \, \mathrm{T}$ magnetic field at room temperature (and then the field was switched off). For the textured sample, the field was applied along the texture (magnetic easy) axis, while for the isotropic sample, the field was directed along an arbitrary in-plane direction. SANS data reduction (correction for background scattering, transmission, detector efficiency) was carried out using the QtiKWS and GRASP software packages~\cite{QtiKWS,graspurl1}.

\section{Results and Discussion}
\label{results}

Figure~\ref{fig1} shows the total SANS cross section of both sintered isotropic and sintered textured Nd-Fe-B at temperatures of $30 \, ^{\circ}\mathrm{C}$ and $450 \, ^{\circ}\mathrm{C}$. While the isotropic sample reveals an isotropic scattering pattern [Fig.~\ref{fig1}(a) and (b)], in agreement with an expected statistically-isotropic grain microstructure on the probed length scale, the textured magnet exhibits a cross-shaped angular anisotropy of $d \Sigma / d \Omega$, persisting well above the Curie temperature of $T_C \cong 283 \, ^{\circ}\mathrm{C}$ up to $1000 \, ^{\circ}\mathrm{C}$ [Fig.~\ref{fig1}(c) and (d), and inset in Fig.~\ref{fig1}(e)]. The scattering pattern of textured Nd-Fe-B is slightly rotated (counterclockwise) around the beam axis by about $8 \, ^{\circ}$, which is due to the misalignment of the texture axis with respect to the horizontal direction. Note that the neutron data below $450 \, ^{\circ}\mathrm{C}$ and at $1000 \, ^{\circ}\mathrm{C}$ were taken in two different sample holders. This explains the different orientation of the angular anisotropy of the $30 \, ^{\circ}\mathrm{C}$ and $450 \, ^{\circ}\mathrm{C}$ data sets [rotated counterclockwise, Fig.~\ref{fig1}(c) and (d)] and the run at $1000 \, ^{\circ}\mathrm{C}$ [rotated clockwise, inset in Fig.~\ref{fig1}(e)]. The observations in Fig.~\ref{fig1} clearly show that the origin of the angular anisotropy in $d \Sigma / d \Omega$ is related to some texture in the nuclear grain microstructure, and not to a specific magnetization distribution or domain structure as concluded earlier for a Nd$_2$Fe$_{14}$B single crystal~\cite{kreyssig09}.

The cross-shaped feature in $d \Sigma / d \Omega$ resembles the nuclear scattering from preferentially-oriented anisometric (e.g., cuboidal) precipitates in metallic alloys (see, e.g., Refs.~\cite{bellet1992,fratzl1993,fratzl1997,deschamps2013} and references therein). Such cuboidal structures were, however, not directly observed in our samples by means of scanning and transmission electron microscopy~\cite{perigo2016no1,perigo2018,perigo2015}.

\begin{figure}[tb!]
\centering
\includegraphics[width=1.0\columnwidth]{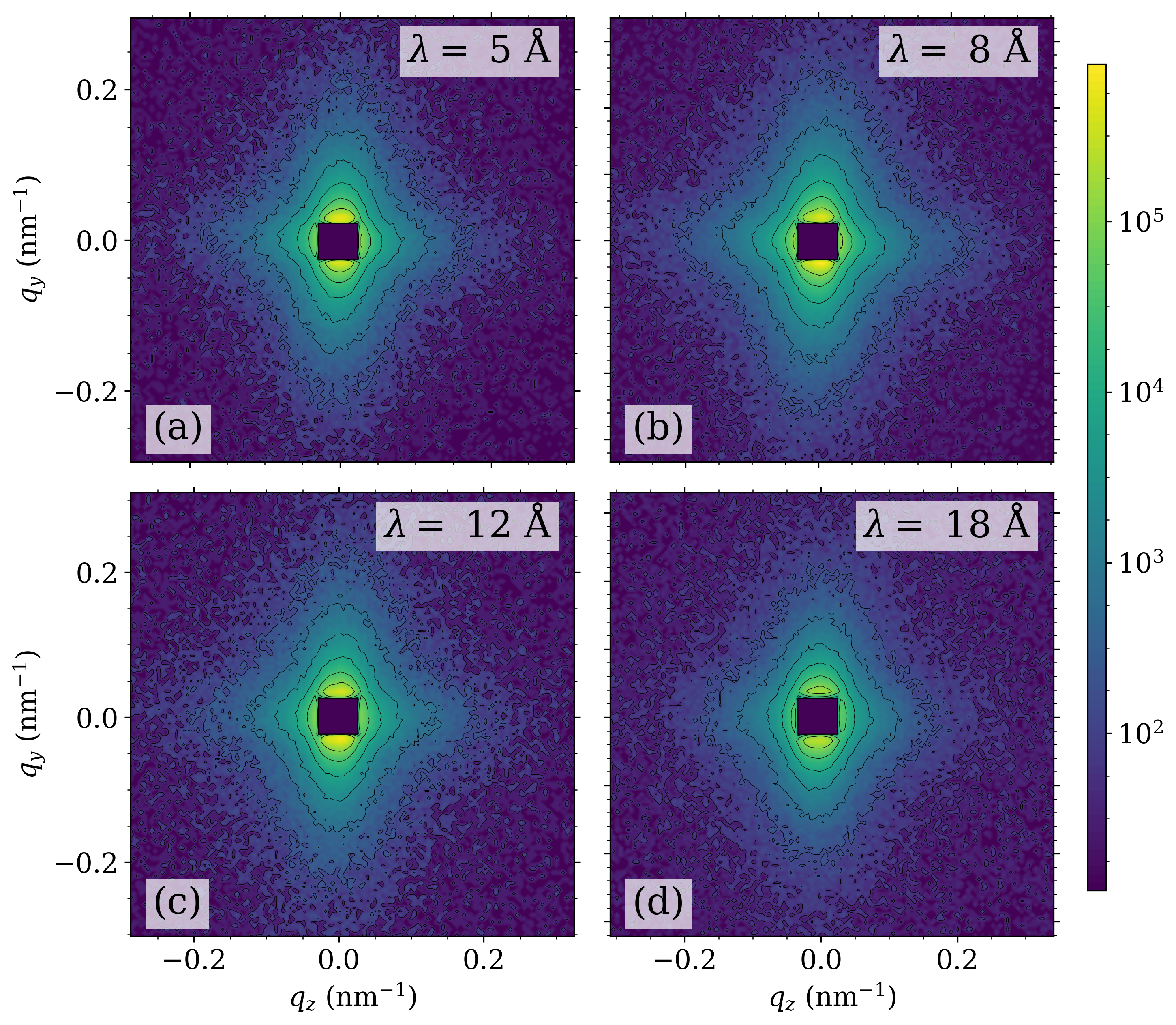}
\caption{\label{fig2} (a)--(d) Total SANS cross section $d \Sigma / d \Omega$ of sintered textured Nd-Fe-B at room temperature as a function of the average neutron wavelength $\lambda$ (logarithmic color scale) (SANS-1, PSI).}
\end{figure}

Figure~\ref{fig2} displays the total SANS cross section $d \Sigma / d \Omega$ of the textured sample at room temperature and for several values of the average neutron wavelength $\lambda$ between $5-18 \, \mathrm{\AA}$. The cross-shaped anisotropy is clearly observed in $d \Sigma / d \Omega$ at all wavelengths. This data set unambiguously proves that double Bragg diffraction~\cite{ohmasa2016} can be excluded as a source for the observed anisotropy, since the maximum value of the lattice-plane spacing in $\mathrm{Nd}_2\mathrm{Fe}_{14}\mathrm{B}$ is $d_{\mathrm{max}} = d_{110} \cong 6.22 \, \mathrm{\AA}$~\cite{pinkerton84}, and no Bragg diffraction from the $\mathrm{Nd}_2\mathrm{Fe}_{14}\mathrm{B}$ unit cell can occur for $\lambda > 2 d_{\mathrm{max}}$; in other words, the origin of the cross-shaped anisotropy in $d \Sigma / d \Omega$ is very likely related to the nuclear small-angle scattering arising from an anisotropic scattering-length density distribution, not Bragg diffraction. Moreover, it is well-known that the multiple-scattering contribution to the single-scattering cross section increases with increasing wavelength. However, since the cross-shaped anisotropy is seen over a rather large range in $\lambda$, we are convinced that the feature is also not related to multiple SANS.

In order to quantitatively characterize the presumed anisometric structure, we have analyzed the experimental nuclear $d \Sigma / d \Omega$ using the NOC software developed by Strunz~\textit{et al.}~\cite{strunz1997,strunz2000,schneider2000,strunz2002,strunz2003,strunz2018}. The microstructural model is based on the superquadrics shape, which in three-dimensional real space is defined by the following inequality~\cite{schneider2000,strunz2002}:
\begin{equation}
\label{superquadrics}
\left( \frac{x-x_0}{R_x} \right)^{2/\sigma} + \left( \frac{y-y_0}{R_y} \right)^{2/\sigma} + \left( \frac{z-z_0}{R_z} \right)^{2/\sigma} \leq 1 ;
\end{equation}
$x_0$, $y_0$, $z_0$ denote the coordinates of the particle center, and $R_x$, $R_y$, $R_z$ are its ``radii'' (or more precisely halves of its size parameters) in $x$, $y$, and $z$-directions, respectively. The parameter $\sigma$, called the morphology parameter by~\textcite{schneider2000}, determines the ``rounding'' of the particle. Obviously, the object is a sphere or an ellipsoid for $\sigma = 1$, and it becomes less rounded when $\sigma$ decreases towards zero. In the limit $\sigma \rightarrow 0$, the object is an exact rectangular parallelepiped. Modeling of SANS single-particle form factors using the superquadrics shape has the advantage that a continuous variety of shapes can be straightforwardly described.

\begin{figure}[tb!]
\centering
\includegraphics[width=1.0\columnwidth]{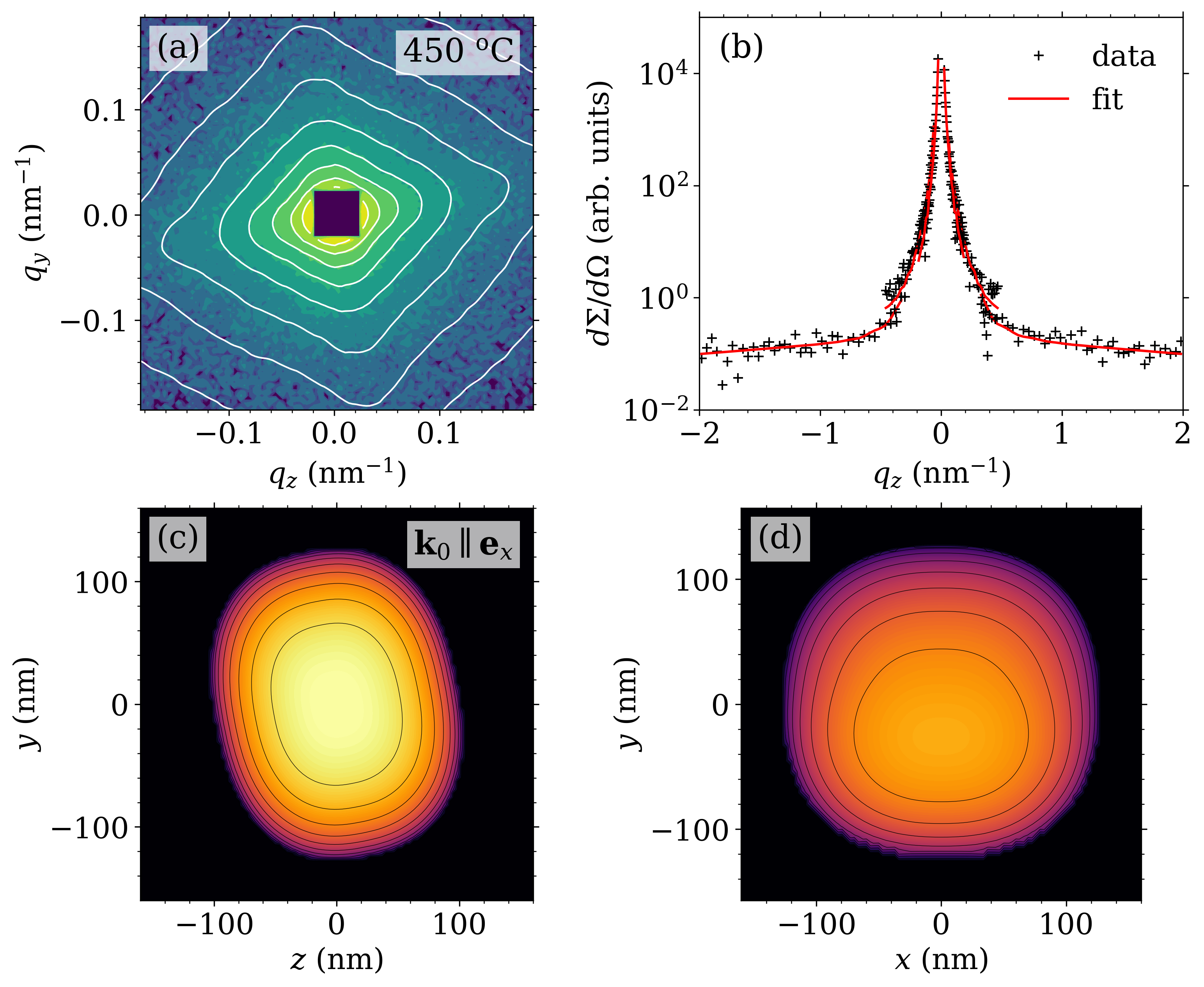}
\caption{\label{fig3} Results of the microstructural SANS modeling based on the superquadrics shape [Eq.~(\ref{superquadrics})]. (a)~Experimental two-dimensional SANS cross section of sintered textured Nd-Fe-B at $450 \, ^{\circ}\mathrm{C}$ (logarithmic color scale). The white equi-intensity contour lines represent the fit based on Eq.~(\ref{superquadrics}) to the experimental data (KWS-1, JCNS). (b)~Horizontal sector average of the data shown in~(a)  (log-linear scale). Solid lines:~fit based on Eq.~(\ref{superquadrics}). (c) and (d)~Model projections of the determined particle shape based on Eq.~(\ref{superquadrics}). The incident neutron beam with wave vector $\mathbf{k}_0$ is along the $\mathbf{e}_x$-direction, while the detector plane is spanned by $\mathbf{e}_y$ and $\mathbf{e}_z$.}
\end{figure}

The result of the above described SANS analysis is summarized in Fig.~\ref{fig3} for the textured sample measured at $450 \, ^{\circ}\mathrm{C}$. The main scattering contribution at all sample-to-detector distances is caused by large structural or compositional inhomogeneities resulting in an anisotropic scattering pattern. As the particles are certainly larger than $100 \, \mathrm{nm}$ (only the asymptotic part of the scattering curve is visible in Fig.~\ref{fig3}), the data do not allow for a reliable determination of the particle size. In fact, from the data in Fig.~\ref{fig3}(a) and (b) we can only estimate a lower bound for the particle dimensions [see Fig.~\ref{fig3}(c) and (d)]. It must also be emphasized that with our microstructural model we cannot distinguish whether the anisotropic scattering is due to some well-defined arrangement of particles or due to larger-scale structural or compositional inhomogeneities. We can only state that the structure is large and that the distribution of interfaces of that structure corresponds to the one of the superquadrics shapes displayed in Fig.~\ref{fig3}(c) and (d). From the fit analysis we obtain for the ratio of the sizes in the $y$ and $z$-direction a value of $1.265 \pm 0.010$ and a morphology parameter of $\sigma = 0.82 \pm 0.01$. The fit error itself is smaller ($\sim 0.2 \, \%$) than the above mentioned ones, and was confirmed by a series of fits with varying initial values of the free fit parameters. The uncertainties in the parameters have been computed from a series of fit results for two different sample-to-detector distances ($8 \, \mathrm{m}$ and $20 \, \mathrm{m}$). The spread of the results for the fitting parameters for the two sample-to-detector distances (i.e., within a rather broad $q$-range) reflects also the possible systematic error due to the imperfection of the used fit model Eq.~(\ref{superquadrics}). The particle (and thus also the scattering pattern) was found to be rotated with respect to the detector counterclockwise around the beam axis ($x$-direction) by $11 \pm 1 \, ^{\circ}$.

\begin{figure}[tb!]
\centering
\includegraphics[width=1.0\columnwidth]{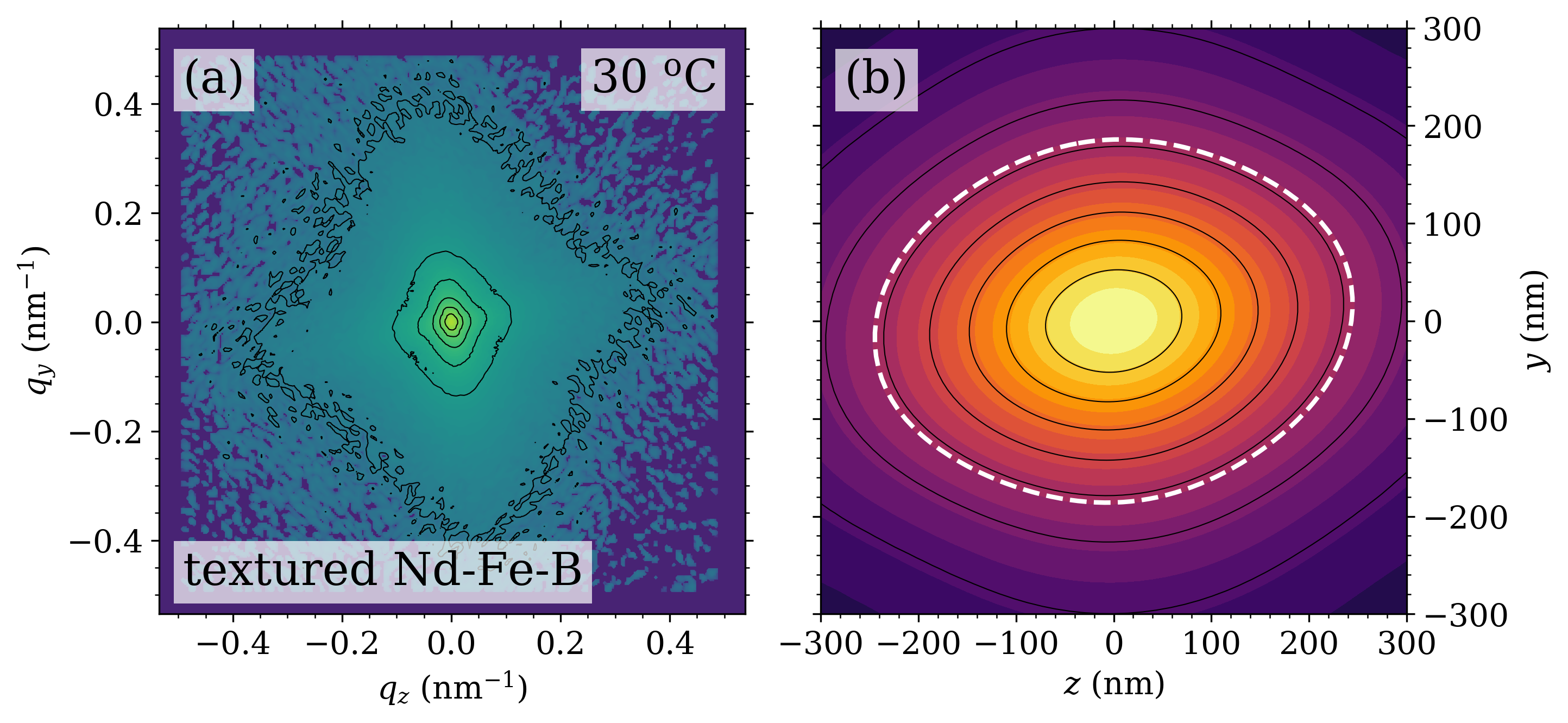}
\caption{\label{fig4} (a)~Room-temperature magnetic SANS cross section $d\Sigma_M / d \Omega$ of textured Nd-Fe-B (logarithmic color scale) (KWS-1, JCNS). The data were obtained by subtracting the purely nuclear neutron data at $450 \, ^{\circ}\mathrm{C}$ [Fig.~\ref{fig1}(d)] from the unpolarized (nuclear and magnetic) SANS cross section at $30 \, ^{\circ}\mathrm{C}$ [Fig.~\ref{fig1}(c)]. (b)~Corresponding two-dimensional magnetic correlation function $c(y,z)$, computed according to Eq.~(\ref{corrfunc}). Dashed white contour line:~$c(y,z) = \exp(-1)$.}
\end{figure}

Assuming that the nuclear SANS is approximately temperature-independent between $450 \, ^{\circ}\mathrm{C}$ and $30 \, ^{\circ}\mathrm{C}$, the room-temperature \textit{magnetic} SANS cross section $d\Sigma_M / d \Omega$ of the textured Nd-Fe-B magnet can be obtained by subtracting the neutron data at $450 \, ^{\circ}\mathrm{C}$ [Fig.~\ref{fig1}(d)] from the data at $30 \, ^{\circ}\mathrm{C}$ [Fig.~\ref{fig1}(c)]. The above assumption is supported by the data displayed in Fig.~\ref{fig1}(e), which reveal that the neutron countrate of the textured sample remains constant above $300 \, ^{\circ}\mathrm{C}$ up to $450 \, ^{\circ}\mathrm{C}$. This observation indicates that in this temperature regime no significant structural transformations appear (e.g., the precipitation of second-phase particles), which in effect could give rise to a temperature-dependent nuclear SANS signal. The in this way obtained $d\Sigma_M / d \Omega$ is displayed in Fig.~\ref{fig4}(a) and exhibits a very similar cross-shaped angular anisotropy as the high-temperature nuclear scattering. This observation suggests that the magnetic moments decorate the anisometric nuclear structure, presumably via the strong spin-orbit coupling in Nd-Fe-B, which is characterized by an uniaxial magnetocrystalline anisotropy of $K_1 \cong 5 \times 10^6 \, \mathrm{J}/\mathrm{m^3}$~\cite{gutfleisch2000}; in other words, the nuclear texture, which is created during the magnet production, gives rise to a room-temperature magnetic texture.

Using the two-dimensional $d\Sigma_M / d \Omega$, the normalized magnetic correlation function $c(y,z)$ can be numerically computed according to~\cite{mettus2015,mettusprm2017}:
\begin{equation}
\label{corrfunc}
c(y,z) = \frac{\int\limits_{-\infty}^{+\infty} \int\limits_{-\infty}^{+\infty} \frac{d\Sigma_M}{d \Omega}(q_y,q_z) \cos(\mathbf{q} \cdot \mathbf{r}) dq_y dq_z}{\int\limits_{-\infty}^{+\infty} \int\limits_{-\infty}^{+\infty} \frac{d\Sigma_M}{d \Omega}(q_y,q_z) dq_y dq_z} ,
\end{equation}
which allows one to estimate the characteristic length scale of the ferromagnetic correlations. The two-dimensional $c(y,z)$ calculated for the textured sample [Fig.~\ref{fig4}(b)] is considerably elongated along the horizontal texture axis. We have estimated a corresponding correlation length $l_C$ as the $\exp(-1)$ decay length [dashed white contour line in Fig.~\ref{fig4}(b)]. The values are $\sim 250 \, \mathrm{nm}$ along the horizontal texture axis and $\sim 190 \, \mathrm{nm}$ perpendicular to it. For comparison, the isotropic sample (data not shown) yields $l_C \sim 220 \, \mathrm{nm}$. These $l_C$-values can be seen as lower bounds for the size of magnetic correlations along the longitudinal and tranversal direction, as only power-law-type magnetic scattering with an exponent of $-5.0$ is observed in this system.

As is well known, the anisotropy of the susceptibility tensor in rare-earth-based metals gives rise to a pronounced and peculiar field dependence of the SANS cross section in the paramagnetic temperature regime ($T > T_C$). Paramagnetic SANS has first been experimentally and theoretically investigated by \textcite{weissm08} on terbium and then later by \textcite{michels2015jpcm} on holmium and gadolinium. With increasing field the SANS cross section increases, usually quite strongly, due to the pronounced magnetic anisotropy of these systems. We emphasize however that our temperature-dependent neutron measurements were all done in zero applied magnetic field, and that the subtraction procedure ($32 – 450 \, ^{\circ}\mathrm{C}$) was therefore also performed on such data. The remanent magnetization of the textured sample above $T_C$ is close to zero, so that the corresponding paramagnetic (field-dependent) SANS under question is presumably small, if not negligible. Magnetic-field-dependent SANS in the paramagnetic state of Nd-Fe-B has not been studied yet and would certainly be of interest.

The origin of the texture in the nuclear grain microstructure remains unknown. Further experiments using very small-angle neutron scattering (VSANS), which would allow us to access momentum-transfers down to about $q_{\mathrm{min}} \cong 0.001 \, \mathrm{nm}^{-1}$ ($\pi / q_{\mathrm{min}} \cong 3 \, \mu\mathrm{m}$), might provide further insights into this question. Moreover, three-dimensional isosurface reconstruction~\cite{hono2009} could provide an answer on the origin of the anisometric structures in sintered Nd-Fe-B.

\section{Conclusion}
\label{con}

Temperature and wavelength-dependent small-angle neutron scattering (SANS) data reveal a cross-shaped angular anisotropy in the SANS cross section of a textured Nd-Fe-B magnet. This anisotropy is shown to be of nuclear origin. Analysis of the two-dimensional purely nuclear SANS data in terms of a microstructural model based on the superquadrics shape form factor suggests that the scattering is due to large ($\gtrsim 100 \, \mathrm{nm}$) anisometric structures. However, it is still unclear whether the associated anisotropic scattering is due to some arrangement of preferentially oriented particles or due to larger-scale structural or compositional inhomogeneities. When cooling the textured magnet from the paramagnetic regime ($T = 450 \, ^{\circ}\mathrm{C}$) through $T_C \cong 283 \, ^{\circ}\mathrm{C}$ to room temperature, we find evidence that the ordered magnetic moments decorate the nuclear texture, with longitudinal and transversal correlation lengths in excess of about $200 \, \mathrm{nm}$.

\section*{Acknowledgements}

This work is based on experiments performed at the Swiss spallation neutron source SINQ, Paul Scherrer Institute, Villigen, Switzerland and at the J\"ulich Centre for Neutron Science (JCNS) at Heinz Maier-Leibnitz Zentrum (MLZ), Garching, Germany. Pavel Strunz acknowledges support from the Czech Academy of Sciences program ``Strategie AV21''.

\bibliographystyle{apsrev4-1}

\end{document}